\documentclass{JHEP3}

\title{Classical Spinning Branes in Curved Backgrounds}
\author{Milovan Vasili\'c, Marko Vojinovi\'c \\ Institute of Physics, P.O.Box
57, 11001 Belgrade,
Serbia \\ E-mail: \email{mvasilic@phy.bg.ac.yu}, \email{vmarko@phy.bg.ac.yu}}

\abstract{The dynamics of a classical branelike object in a curved background
is derived from the covariant stress-energy conservation of the brane matter.
The world sheet equations and boundary conditions are obtained in the
pole-dipole approximation, where nontrivial brane thickness gives rise to its
intrinsic angular momentum. It is shown that intrinsic angular momentum
couples to both, the background curvature and the brane orbital degrees of
freedom. The whole procedure is manifestly covariant with respect to spacetime
diffeomorphisms and world sheet reparametrizations. In addition, two extra
gauge symmetries are discovered and utilized. The examples of the point
particle and the string in $4$ spacetime dimensions are analyzed in more
detail. A particular attention is paid to the Nambu-Goto string with massive
spinning particles attached to its ends.}

\keywords{p-branes, Classical Theories of Gravity}

\preprint{}

\usepackage{amsmath}

\newcommand{\itGamma}{{\mathit{\Gamma}}}
\newcommand{\del}{\partial}

\newcommand{\cM}{{\cal M}}
\newcommand{\cO}{{\cal O}}
\newcommand{\ds}{\displaystyle}
\newcommand{\ts}{\textstyle}
\newcommand{\lc}{\varepsilon}
\newcommand{\h}{h}
\newcommand{\diag}{\mathop{\rm diag}\nolimits}
\newcommand{\orto}{{\scriptscriptstyle\perp}}
\newcommand{\para}{\scriptscriptstyle\parallel}
\newcommand{\Pp}{ {P_{\para}}\vphantom{P} }
\newcommand{\Ppp}{ {p_{\para}}\vphantom{p} }
\newcommand{\Pn}{ {P_{\orto}}\vphantom{P} }
\newcommand{\Pnn}{ {p_{\orto}}\vphantom{p} }

\begin{document}

\section{\label{IntroductionSection}Introduction}

The interest in studying extended objects in high energy physics began with
the observation that meson resonances could be viewed as rotating relativistic
strings. This model provided a successful explanation of Regge trajectories
and Veneziano amplitudes. In later development of the idea, relativistic
strings have been promoted to elementary building blocks of the known matter,
and as such extensively studied.

A parallel line of research treated strings as linelike kink solutions in a
field theory. Such is, for example, the Nielsen-Olesen vortex line solution of
a Higgs type scalar electrodynamics \cite{Nielsen1973}. The idea behind this
approach is to try and describe bound states of quarks as flux tube solutions
of the Standard model.

Whatever idea guides one to explore strings, or more generally branes, the
general form of their classical dynamics may be needed. In this paper, we
shall be concerned with the classical branelike kink configurations in an
arbitrary Riemannian spacetime. We shall not specify the type of matter the
branes are made of, but merely assume that such kink configurations exist. For
simplicity, the dynamics of spacetime geometry is assumed to be that of
general relativity. In this setting, the stress-energy tensor of matter fields
is symmetric, $T^{\mu\nu} = T^{\nu\mu}$, and covariantly conserved,
\begin{equation} \label{jna1}
\nabla_{\nu}T^{\mu\nu}=0 \, .
\end{equation}

The covariant conservation law of the stress-energy tensor $T^{\mu\nu}$ is the
starting point in our analysis of brane dynamics in curved spacetimes. The
method we use is a generalization of the Mathisson-Papapetrou method for
pointlike matter \cite{Mathisson1937, Papapetrou1951}. It has already been
exploited in ref.\ \cite{Vasilic2006} for the study of stringlike objects in
the lowest (single-pole) approximation. There, the world sheet effective
equations of motion are obtained from the conservation equations (\ref{jna1})
in the limit of an infinitely thin string. In this paper, we extend the
analysis to the next level of approximation --- the pole-dipole approximation.
In this approximation, a nonzero thickness of the brane is taken into
consideration. As we shall see, the additional degrees of freedom that thus
appear account for the internal angular momentum of the brane.

The motivation for studying classical branelike matter in curved backgrounds
is threefold. First, we believe it is useful to have fully covariant
description of a classical $p$-brane with intrinsic angular momentum. We
restrict ourselves to Riemannian spacetimes, but the analysis can be extended
to include torsion. The dimensions of the brane ($p$) and the
spacetime ($D$) remain arbitrary. Second, we find it interesting to try and
extend the known Nambu-Goto string (described by the tension alone) to allow a
nontrivial intrinsic angular momentum. A simple model of the kind could,
at least, give
us a clue to what kind of dynamics one could expect from the spinning string.
Finally, our basic motivation for this work is proper preparation for treating
strings in spacetimes with torsion. In the existing literature, the influence
of torsion has been studied in the case of pointlike matter only. It has been
suggested that the consistent treatment of the problem demands pole-dipole
approximation \cite{Yasskin1980, Nomura1991}. Naturally, we expect the same in
the case of strings and higher branes.

The new results obtained in this paper can be summarized as follows. First,
the Mathisson-Papapetrou method has been generalized for the treatment of
higher branes in curved backgrounds. We have refined the method by developing
a manifestly covariant decomposition of the stress-energy tensor into a series
of $\delta$-function derivatives. It has been shown that a truncation of the
series is covariant with respect to both, spacetime diffeomorphisms and world
sheet reparametrizations. In addition, two extra gauge symmetries are
discovered and analyzed. The extra gauge fixing has been shown to define
centre of mass of pointlike matter, and its generalization to central
surface of mass in the case of branelike matter. Second, the fully covariant
world sheet equations and boundary conditions of a $p$-brane in a
$D$-dimensional Riemannian spacetime have been obtained in the pole-dipole
approximation. The general brane dynamics turns out to depend on the effective
$(p+1)$-dimensional stress-energy tensor of the brane, and $(p+1)$-dimensional
currents corresponding to its internal angular momentum. We have utilized the
discovered extra gauge freedom to show that charges corresponding to internal
boosts can be gauged away. Finally, particles and strings in $4$-dimensional
spacetime have been analyzed in more detail. It has been shown that the
Nambu-Goto string can be generalized to include spinning matter on its ends,
thereby providing a better model for meson resonances. In the case of
pointlike matter, the known Papapetrou results are reproduced
\cite{Papapetrou1951}.

The layout of the paper is as follows. In section\ \ref{MultipoleSection}, we
define covariant decomposition of the stress-energy tensor as a series of
$\delta$-function derivatives. We demonstrate the invariance with respect to
spacetime diffeomorphisms, world sheet reparametrizations and two additional
gauge transformations. The truncation of the series is shown to respect
general covariance. In section\ \ref{EquationsSection}, the world sheet equations
and boundary conditions are derived in pole-dipole approximation. This is done
by neglecting all but the first two terms in the decomposition of the
stress-energy tensor, and using it in the conservation equations (\ref{jna1}).
In section\ \ref{InterpretationSection}, we analyze the symmetry properties of
the
free coefficients of our world sheet equations and boundary conditions. We
identify the effective $(p+1)$-dimensional stress-energy tensor of the brane,
and $(p+1)$-dimensional currents associated with its intrinsic angular
momentum. These are the only free parameters that affect the brane dynamics in
the pole-dipole approximation. In section\ \ref{ExamplesSection}, the examples of
pointlike and stringlike matter are considered. In particular, the
Nambu-Goto string is generalized to allow spinning matter on its ends. Section
\ref{ConclusionSection} is devoted to concluding remarks.

Our conventions are the same as in ref.\ \cite{Vasilic2006}, with the
exception
of the metric signature. Greek indices $\mu,\nu,\dots$ are the spacetime
indices, and run over $0,1,\dots,D-1$. Latin indices $a,b,\dots$ are the world
sheet indices and run over $0,1,\dots,p$. The Latin indices $i,j,\dots$ refer
to the world sheet boundary and take values $0,1,\dots,p-1$. The coordinates
of spacetime, world sheet and world sheet boundary are denoted by $x^{\mu}$,
$\xi^a$ and $\lambda^i$, respectively. The corresponding metric tensors are
denoted by $g_{\mu\nu}(x)$, $\gamma_{ab}(\xi)$ and $\h_{ij}(\lambda)$. The
signature convention is defined by $\diag(-,+,\dots,+)$, and the indices are
raised by the inverse metrics $g^{\mu\nu}$, $\gamma^{ab}$ and $\h^{ij}$.

\section{\label{MultipoleSection}Multipole expansion of the stress-energy
tensor}

A $p$-brane is an extended $p$-dimensional object whose trajectory is a
$(p+1)$-dimensional world sheet, commonly denoted by $\cM$. In this paper, we
shall be concerned with material objects shaped to resemble a $p$-brane. If
this is the case, all but the first couple of terms in the multipole expansion
around a suitably chosen $(p+1)$-dimensional surface can be neglected.
Retaining the first two terms defines the so called pole-dipole approximation.

Let us begin with the introduction of a $(p+1)$-dimensional surface
$x^{\mu}=z^{\mu}(\xi)$ in $D$-dimensional spacetime, where $\xi^a$ are the
surface coordinates. We shall assume that the surface is \emph{everywhere
regular}, and the coordinates $\xi^a$ well defined. We shall consider only
time infinite brane trajectories. This means that every spatial section of the
spacetime has nonempty intersection with the world sheet. As for the
intersection itself, it is supposed to be of finite length. Thus, only closed
or finite open branes are considered. The world sheet boundary $\del\cM$ is
parametrized by $p$ coordinates $\lambda^i$.

In what follows, we shall frequently use the notion of the world sheet
coordinate vectors
$$
u_a^{\mu} = \frac{\del z^{\mu}}{\del \xi^a} \, ,
$$
and the world sheet induced metric tensor
$$
\gamma_{ab} = g_{\mu\nu} u_a^{\mu} u_b^{\nu} \, .
$$
The induced metric is assumed to be nondegenerate, $\gamma \equiv
\det(\gamma_{ab}) \neq 0$, and of Minkowski signature. With this assumption,
each point on the world sheet accommodates a timelike tangent vector. This is
how the notion of the timelike curve is generalized to a $(p+1)$-dimensional
case.

Now, we are ready to expand the stress-energy tensor into a $\delta$-function
series around the surface $x^{\mu} = z^{\mu}(\xi)$. Generalizing the results
of ref.\ \cite{Vasilic2006}, where this expansion has been used in the
single-pole approximation, we define
\begin{equation} \label{jna2}
T^{\mu\nu}(x) = \int d^{p+1}\xi \sqrt{-\gamma} \left[ b^{\mu\nu}(\xi)
\frac{\delta^{(D)}(x-z)}{\sqrt{-g}}
+ b^{\mu\nu\rho}(\xi) \nabla_{\rho}
\frac{\delta^{(D)}(x-z)}{\sqrt{-g}} + \dots \right],
\end{equation}
where $\nabla_{\rho}$ stands for the Riemannian covariant derivative, and
$g(x)$ is the determinant of the target space metric $g_{\mu\nu}(x)$.

The decomposition (\ref{jna2}) is suitable for treating matter which is well
localized around the brane $x^{\mu}=z^{\mu}(\xi)$. In fact, the stress-energy
tensor $T^{\mu\nu}(x)$ must drop exponentially to zero as we move away from
the brane if we want the series (\ref{jna2}) to be well defined. If this is
the case, each coefficient $b^{\mu\nu\rho_1\dots\rho_n}$ gets smaller as $n$
gets larger. In the lowest, single-pole approximation, all $b$'s except the
first are neglected, and we end up with the manifestly covariant expression
analyzed in ref.\ \cite{Vasilic2006}. In this paper, we extend the analysis to
the pole-dipole approximation, defined by neglecting all but the first two
$b$-coefficients.

\subsection{Diffeomorphism invariance}

The series (\ref{jna2}) can, in general, be truncated at any level. As opposed
to the single-pole approximation, however, the general truncation turns out
not to be manifestly covariant. Indeed, the transformation properties of the
$b$-coefficients, as derived from the known transformation law of the
stress-energy tensor $T^{\mu\nu}$, show that $b$'s are not tensors. This
leaves us with two tasks to be accomplished. The first is to show that the
general truncation of the series (\ref{jna2}) is diffeomorphism invariant. The
second is to find a manifestly covariant form of the truncated expression.

Let us start with transformation properties of the $b$-coefficients. First, we
define the scalar functional
\begin{equation} \label{jna3}
T[f] \equiv \int d^Dx \sqrt{-g}\, T^{\mu\nu}(x) f_{\mu\nu}(x) \, ,
\end{equation}
where $f_{\mu\nu}(x)$ is an arbitrary tensor field with compact support. The
decomposition (\ref{jna2}) then yields
\begin{equation} \label{jna4}
T[f] = \int d^{p+1}\xi \sqrt{-\gamma} \Big[ I_0\left( b_0 f \right) + I_1
\left( b_1 f \right) + \dots \Big] \, ,
\end{equation}
where
\begin{equation} \label{jna5}
I_n\left( b_n f \right) = \int d^Dx \sqrt{-g} \, b^{\mu\nu\rho_1\dots\rho_n}
f_{\mu\nu} \nabla_{\rho_1} \dots \nabla_{\rho_n}
\frac{\delta^{(D)}(x-z)}{\sqrt{-g}} \, .
\end{equation}
We can now make use of the compact support of the arbitrary functions
$f_{\mu\nu}(x)$ to perform a series of partial integrations in (\ref{jna5}).
This leads to
\begin{equation} \label{jna6}
I_n\left( b_n f \right) = (-1)^n \nabla_{\rho_n} \dots \nabla_{\rho_1}
b^{\mu\nu\rho_1\dots\rho_n} f_{\mu\nu} \Big|_{x=z} \, .
\end{equation}
In this expression, the action of the covariant derivative $\nabla_{\rho}$ on
the $b$-coefficients is defined formally, by treating $b$'s as tensors with no
$x$ dependence. Thus, the expression $\nabla b_n$ contains only $\itGamma b_n$
terms in accordance with the index structure of the $b_n$ coefficient (e.g.
$\nabla_{\rho} b^{\mu\nu}(\xi) = {\itGamma^{\mu}}_{\lambda\rho} (x)
b^{\lambda\nu}(\xi) + {\itGamma^{\nu}}_{\lambda\rho}(x) b^{\mu\lambda}(\xi)$).
Now, we perform differentiations in (\ref{jna6}). The result can symbolically
be written as
\begin{eqnarray} \nonumber
I_0 (b_0 f) & = & b_0 f(z), \\ \nonumber
I_1 (b_1 f) & = & - \left[ (\nabla b_1) f + b_1 \nabla f \right]_{x=z} \, , \\
\nonumber
I_2 (b_2 f) & = & \left[ (\nabla^2 b_2) f + 2(\nabla b_2) \nabla f + b_2
\nabla^2 f \right]_{x=z} \, , \ \dots
\end{eqnarray}
Using this in the decomposition (\ref{jna4}), we can regroup the additive
terms to obtain
\begin{equation} \label{jna7}
\ds T[f]= \int d^{p+1}\xi \sqrt{-\gamma} \Big[ B^{\mu\nu}f_{\mu\nu}(z) +
B^{\mu\nu\rho}f_{\mu\nu;\rho}(z) + B^{\mu\nu\rho\lambda}
f_{\mu\nu;\rho\lambda}(z) +\dots \Big].
\end{equation}
Here, $f_{\mu\nu;\rho_1\dots\rho_n}(z)$ stands for $\nabla_{\rho_n} \dots
\nabla_{\rho_1} f_{\mu\nu}$ evaluated at $x=z(\xi)$, and the coefficients
$B(\xi)$ have the general structure
\begin{subequations} \label{jna8}
\begin{equation} \label{jna8a}
\begin{array}{lcccccccccl}
B_0 & =      & b_0 & - & \nabla b_1 & + & \nabla^2 b_2 & - & \nabla^3 b_3  & +
& \dots \,  , \\
B_1 & =      &     & - & b_1        & + & 2\nabla b_2  & - & 3\nabla^2 b_3 & +
& \dots  \, , \\
B_2 & =      &     &   &            &   & b_2          & - & 3 \nabla b_3  & +
& \dots  \, ,\ \dots \\
\end{array}
\end{equation}
We see that the system of equations (\ref{jna8a}) can be solved for $b$'s.
Symbolically,
\begin{equation} \label{jna8b}
\begin{array}{lcccccccccl}
b_0 & =      & B_0 & - & \nabla B_1 & + & \nabla^2 B_2 & - & \nabla^3 B_3  & +
& \dots  \, , \\
b_1 & =      &     & - & B_1        & + & 2\nabla B_2  & - & 3\nabla^2 B_3 & +
& \dots  \, , \\
b_2 & =      &     &   &            &   & B_2          & - & 3 \nabla B_3  & +
& \dots  \, , \ \dots \\
\end{array}
\end{equation}
\end{subequations}

The obtained results lead us to two important conclusions. First,
\begin{itemize}
\item $B$-coefficients are tensors with respect to spacetime
diffeomorphisms.
\end{itemize}
This is a consequence of the fact that $T[f]$ in (\ref{jna7})
is a scalar functional for any choice of the tensor field $f_{\mu\nu}(x)$. The
corresponding transformation law reads
\begin{equation} \label{jna9}
{B'}^{\mu_1\dots\mu_n} = \left( \frac{\del {x'}^{\mu_1}}{\del x^{\nu_1}} \dots
\frac{\del {x'}^{\mu_n}}{\del x^{\nu_n}} \right)_{x=z} B^{\nu_1\dots\nu_n}  \,
.
\end{equation}
The transformation properties of the $b$-coefficients are derived from
(\ref{jna8b}), and do not have tensorial character. Second,
\begin{itemize}
\item truncation of the series (\ref{jna2}) at any level is a
covariant operation.
\end{itemize}
Indeed, if all the $b$'s of the order $n$ and higher are
put to zero ($b_n=b_{n+1}=\dots=0$), the corresponding $B$'s will also vanish
($B_n=B_{n+1}=\dots=0$), as is seen from (\ref{jna8a}). Being tensors, the
zero $B$'s will remain to be zero in any reference frame
($B_n'=B_{n+1}'=\dots=0$), and according to (\ref{jna8b}) so will the
corresponding $b$'s ($b_n'=b_{n+1}'=\dots=0$). Thus, the truncation is
diffeomorphism invariant.

Let us consider two simple examples. The \emph{single-pole} approximation
is defined by retaining only the
leading term $b_0$, while $b_1=b_2=\dots=0$. The equations (\ref{jna8a}) then
give
$$
b^{\mu\nu} = B^{\mu\nu} \, ,
$$
which means that $b^{\mu\nu}$, in the single-pole approximation, transforms as
a tensor. The stress-energy tensor has a manifestly covariant form
$$
T^{\mu\nu}(x) = \int d^{p+1}\xi \sqrt{-\gamma} b^{\mu\nu}
\frac{\delta^{(D)}(x-z)}{\sqrt{-g}} \, .
$$
In the \emph{pole-dipole} approximation, the first two coefficients $b_0$
and $b_1$ are retained, while the remaining $b_2,b_3,\dots$ are put to zero.
The system of equations (\ref{jna8a}) reduces to
\begin{equation} \label{jna10}
B^{\mu\nu}=b^{\mu\nu} - \nabla_{\rho}b^{\mu\nu\rho}, \qquad B^{\mu\nu\rho} = -
b^{\mu\nu\rho} \, ,
\end{equation}
where $\nabla_{\rho} b^{\mu\nu\rho} \equiv {\itGamma^{\mu}}_{\lambda\rho}
b^{\lambda\nu\rho} + {\itGamma^{\nu}}_{\lambda\rho} b^{\mu\lambda\rho} +
{\itGamma^{\rho}}_{\lambda\rho} b^{\mu\nu\lambda}$. We see that the
coefficient $b^{\mu\nu\rho}$ transforms as a tensor in this approximation,
while $b^{\mu\nu}$ does not. The stress-energy tensor is rewritten in a
manifestly covariant form
\begin{equation} \label{jna11}
\ds T^{\mu\nu}(x) = \int d^{p+1}\xi \sqrt{-\gamma} \left[ B^{\mu\nu}
\frac{\delta^{(D)}(x-z)}{\sqrt{-g}}
- \nabla_{\rho} \left( B^{\mu\nu\rho}
\frac{\delta^{(D)}(x-z)}{\sqrt{-g}} \right) \right] \,  .
\end{equation}

In this form, the decomposition of the stress-energy tensor is manifestly
covariant with respect to both,
spacetime diffeomorphisms and world sheet reparametrizations. The
corresponding transformation properties are summarized as follows:
\begin{center}
\begin{tabular}{|c|c|c|c|c|} \hline
 & $B^{\mu\nu}(\xi)$ & $B^{\mu\nu\rho}(\xi)$ & $\gamma_{ab}(\xi)$ &
$g_{\mu\nu}(x)$ \\ \hline
spacetime & tensor & tensor & scalar & tensor \\ \hline
world sheet & scalar & scalar & tensor & scalar \\ \hline
\end{tabular}
\end{center}

\subsection{Extra symmetry 1}

In this subsection, we shall demonstrate the appearance of an additional gauge
transformation that leaves the stress-energy tensor invariant. To this end,
note that each term in the decomposition (\ref{jna2}) basically contains
$D-p-1$ $\delta$-functions, which are used to model a $p$-brane in a
$D$-dimensional spacetime. The extra $p+1$ $\delta$-functions and extra $p+1$
integrations are introduced only to covariantize the expressions. This
observation leads us to conclude that there are redundant $b$-coefficients in
(\ref{jna2}). In particular, the derivatives parallel to the world sheet are
integrated out, as they should, considering the fact that matter is not
localized along the brane. As a consequence, the parallel components of
$b^{\mu\nu\rho}$ coefficients are expected to dissapear from the decomposition
(\ref{jna2}). To check this, let us define the transformation law of the form
\begin{subequations} \label{jna12}
\begin{equation} \label{jna12a}
\delta_1 b^{\mu\nu\rho} = \epsilon^{\mu\nu a} u_a^{\rho} \, ,
\end{equation}
where $\epsilon^{\mu\nu a}(\xi) = \epsilon^{\nu\mu a}(\xi)$ are free
parameters. Using (\ref{jna12a}) to calculate the variation of the functional
$T[f]$, we find that the invariance of the stress-energy tensor requires an
additional transformation of the $B^{\mu\nu}$ coefficients. Precisely,
\begin{equation} \label{jna12b}
\delta_1 B^{\mu\nu} = -\nabla_a \epsilon^{\mu\nu a} \, ,
\end{equation}
\end{subequations}
where $\nabla_a$ stands for the total covariant derivative, defined in the
appendix. In fact, the transformation law (\ref{jna12})
defines a symmetry of the stress-energy tensor only if the boundary terms are
missing. Indeed, the variation of the functional $T[f]$ under (\ref{jna12})
has the form
$$
T'[f] = T[f] - \int_{\del\cM} d^p\lambda \sqrt{-\h} \, n_a\epsilon^{\mu\nu
a}f_{\mu\nu} \, ,
$$
where $\h_{ij}(\lambda)$ is the induced metric on $\del\cM$, and
$n^a(\lambda)$ is the unit boundary normal (see the appendix). To have the
full invariance, the parameters $\epsilon^{\mu\nu a}$ are required to obey the
boundary conditions
\begin{equation} \label{jna13}
n_a \epsilon^{\mu\nu a} \Big|_{\del\cM} =0 \, .
\end{equation}
The transformation rule (\ref{jna12}),
with parameters constrained by (\ref{jna13}), defines the
\emph{extra symmetry}~$1$ of the brane dynamics.

Now we see that parallel components of the $b^{\mu\nu\rho}$ coefficients are
indeed pure gauge,
$$
\delta_1 (b^{\mu\nu\rho} u^a_{\rho}) = \epsilon^{\mu\nu a} \, .
$$
They can be gauged away everywhere except on the boundary, where the
parameters $\epsilon^{\mu\nu a}$ are not free. As a consequence, the theory
will contain some peculiar degrees of freedom, which live exclusively on the
boundary, and do not appear in the world sheet equations. In the next section,
we shall clarify their physical meaning.

\subsection{Extra symmetry 2}

The expansion of the stress-energy tensor into a $\delta$-function series
(\ref{jna2}) has been performed with an arbitrary choice of the surface
$x^{\mu}=z^{\mu}(\xi)$. If we use another surface, let us say $x^{\mu} =
z'^{\mu}(\xi)$, the coefficients $b^{\mu\nu}$, $b^{\mu\nu\rho}$, $\dots$ will
change to $b'^{\mu\nu}$, $b'^{\mu\nu\rho}$, $\dots$, while leaving the
stress-energy tensor invariant. The transformation law of the
$b$-coefficients, generated by the replacement $z^{\mu}\to z'^{\mu}$, defines
the gauge symmetry that we shall call \emph{extra symmetry}~2.

The extra symmetry 2 is an exact symmetry of the full expansion (\ref{jna2}),
but only approximate symmetry of the truncated series (\ref{jna11}). This is
because the condition $b_2=b_3=\dots=0$ is a gauge condition that fixes the
choice of the surface $x^{\mu}=z^{\mu}(\xi)$. Indeed, if the surface is chosen
to lie outside the region where matter is localized, the higher $b$'s will
give a substantial contribution to the series, no matter how thin the brane
is. The best we can do is to keep the surface $x^{\mu}=z^{\mu}(\xi)$ inside
the localized matter. Then, we can assume the following hierarchy of the
$b$-coefficients:
$$
b^{\mu\nu} = \cO_0, \quad b^{\mu\nu\rho} = \cO_1 \, , \quad
b^{\mu\nu\rho\lambda}=\cO_2 \, , \ \dots \, ,
$$
where $\cO_n$ stands for the order of smallness. The truncation of the $n$-th
order is defined as an approximation in which the $\cO_{n+1}$ and higher terms
are neglected. In this approximation, the parameters of the extra symmetry 2,
as defined by
\begin{equation} \label{jna14}
z'^{\mu}(\xi) = z^{\mu}(\xi) + \epsilon^{\mu}(\xi) \, ,
\end{equation}
are constrained by the requirement
\begin{equation} \label{jna15}
b'_{n+1} = \cO_{n+1} \, .
\end{equation}
The transformation law (\ref{jna14}) generates the corresponding
transformation of the $b$-co\-eff\-i\-ci\-ents. It is shown to have the
general
form
\begin{equation} \label{jna16}
\begin{array}{lcl}
b_0' & = & b_0 + b_0 \epsilon + b_1 \epsilon + \dots \, , \\
b_1' & = & b_1 + b_0 \epsilon + b_1 \epsilon + \dots \, , \\
b_2' & = & b_2 + b_1 \epsilon + b_2 \epsilon + \dots \, , \ \dots \\
\end{array}
\end{equation}
In both, single-pole and pole-dipole approximations, the transformation
(\ref{jna16}) and the constraint (\ref{jna15}) imply
\begin{equation} \label{jna17}
\epsilon^{\mu} = \cO_1  \, .
\end{equation}
This condition ensures that the order of truncation is not spoiled by the
action of extra symmetry 2.

The transformation rule (\ref{jna16}) can be rewritten in terms of
$B$-coefficients. Discarding contributions of the order $\cO_2$ and higher, we
obtain
\begin{equation} \label{jna18}
\begin{array}{lcl}
\delta_2 B^{\mu\nu} & = & - B^{\mu\nu} u^a_{\rho} \nabla_a \epsilon^{\rho} - 2
B^{\lambda(\mu}{\itGamma^{\nu)}}_{\lambda\rho}\epsilon^{\rho} \, , \\
\delta_2 B^{\mu\nu\rho} & = & - B^{\mu\nu}\epsilon^{\rho} \, . \\
\end{array}
\end{equation}
The equations (\ref{jna18}) and (\ref{jna14}) define the extra symmetry 2 in
the pole-dipole approximation.

Three remarks are in order. First, notice that the parameter $\epsilon^{\mu}$,
as defined by (\ref{jna14}), is a spacetime vector. Then, the explicit
presence of the connection in the transformation law (\ref{jna18}) seems to
contradict the tensorial character of the $B$-coefficients. In fact, there is
no contradiction. The transformation law of the $B$-coefficients under
spacetime diffeomorphisms is given by (\ref{jna9}). We see that all
$x$-dependent terms are evaluated on the surface $x^{\mu}=z^{\mu}(\xi)$. When
the surface is changed by (\ref{jna14}), the new coefficients are given by
(\ref{jna18}). Their transformation law under diffeomorphisms is shown to have
the same form as in (\ref{jna9}), the only difference being that the
$x$-dependent terms are now evaluated on the new surface, $x^{\mu} =
z'^{\mu}(\xi)$.

The second remark concerns the single-pole approximation. We have seen that
the invariance of every truncation implies the constraint
$\epsilon^{\mu}=\cO_1$. Using this in the single-pole approximation, which is
defined by neglecting $\cO_1$ terms, we obtain $\delta_2 z^{\mu}=0$, $\delta_2
B^{\mu\nu}=0$. Thus, the extra symmetry 2 in the single-pole approximation is
trivial. This is a consequence of the fact that single-pole branes are
infinitely thin, which leaves no freedom for the choice of $z^{\mu}(\xi)$.

Finally, let us observe that fixing the gauge of extra symmetry 2 defines what
could be called \emph{central surface of mass} distribution for our
localized matter. In the particle case ($0$-brane), this coincides with the
usual notion of the centre of mass. We shall see in
section~\ref{InterpretationSection} how a proper definition of the central
surface of mass simplifies the world sheet equations, and helps us to
interpret free parameters of the theory.

\section{\label{EquationsSection}World sheet equations}

In this section we shall analyze the stress-energy conservation equation
(\ref{jna1}) in the pole-dipole approximation. We first define an arbitrary
vector field $f_{\mu}(x)$ of \emph{compact support}, and rewrite the
equation (\ref{jna1}) in the convenient form
\begin{equation} \label{jna19}
\int d^Dx \sqrt{-g} f_{\mu} \nabla_{\nu} T^{\mu\nu} =0 \, , \qquad \forall
f_{\mu}(x) \, .
\end{equation}
Now, we use the decomposition (\ref{jna11}) of the stress-energy tensor. Owing
to the compact support of $f_{\mu}(x)$, we are allowed to change the order of
integrations, and to drop surface terms. Thus, we arrive at
\begin{equation} \label{jna20}
\int d^{p+1}\xi \sqrt{-\gamma} \left( B^{\mu\nu} f_{\mu;\nu} +
B^{\mu\nu\rho}f_{\mu;\nu\rho} \right) =0 \, ,
\end{equation}
where $f_{\mu;\nu}\equiv (\nabla_{\nu}f_{\mu})_{x=z}$, $f_{\mu;\nu\rho} \equiv
(\nabla_{\rho}\nabla_{\nu} f_{\mu})_{x=z}$. The fact that this equation holds
for every $f_{\mu}(x)$ puts some constraints on the coefficients $B^{\mu\nu}$
and $B^{\mu\nu\rho}$. To find these, we decompose the derivatives of the
vector field $f_{\mu}(x)$ into components orthogonal and parallel to the
surface $x^{\mu}=z^{\mu}(\xi)$:
\begin{subequations} \label{jna21}
\begin{equation} \label{jna21a}
f_{\mu;\lambda} = f^{\orto}_{\mu\lambda} + u^a_{\lambda} \nabla_a f_{\mu} \, ,
\end{equation}
\begin{equation} \label{jna21b}
f_{\mu;(\lambda\rho)} = f^{\orto}_{\mu\lambda\rho} + 2 f^{\orto}_{\mu(\lambda
a} u^a_{\rho)} + f_{\mu ab} u^a_{\lambda} u^b_{\rho} \, ,
\end{equation}
\begin{equation} \label{jna21c}
f_{\mu;[\lambda\rho]} = \frac{1}{2} {R^{\sigma}}_{\mu\lambda\rho} f_{\sigma}
\,
.
\end{equation}
\end{subequations}
Here, the orthogonal and parallel components are obtained by using the
projectors
\begin{equation} \label{jna22}
\Pn^{\mu}_{\nu} = \delta^{\mu}_{\nu} - u_a^{\mu}u^a_{\nu}, \qquad
\Pp^{\mu}_{\nu} = u_a^{\mu} u^a_{\nu}  \, .
\end{equation}
More precisely, $f^{\orto}_{\mu\lambda} = \Pn^{\sigma}_{\lambda}
f_{\mu;\sigma}$,
$f^{\orto}_{\mu\lambda\rho} = \Pn^{\sigma}_{\lambda} \Pn^{\nu}_{\rho}
f_{\mu;(\sigma\nu)}$, $f^{\orto}_{\mu\lambda a} = \Pn^{\sigma}_{\lambda}
u_a^{\nu} f_{\mu;(\sigma\nu)}$ and $f_{\mu ab} = u_a^{\sigma} u_b^{\nu}
f_{\mu;(\sigma\nu)}$. Direct calculation yields
\begin{equation} \label{jna23}
\begin{array}{lcl}
f_{\mu ab} & = & \nabla_{(a} \nabla_{b)} f_{\mu} - (\nabla_a u_b^{\nu})
f^{\orto}_{\mu\nu} \, , \\
f^{\orto}_{\mu\rho a} & = & \ds \Pn^{\nu}_{\rho} \nabla_a f^{\orto}_{\mu\nu} +
(\nabla_a u^b_{\rho}) \nabla_b f_{\mu} + \frac{1}{2} \Pn^{\lambda}_{\rho}
u_a^{\nu} {R^{\sigma}}_{\mu\nu\lambda} f_{\sigma} \, , \\
\end{array}
\end{equation}
which tells us that the only independent components on the surface
$x^{\mu}=z^{\mu}(\xi)$ are $f_{\mu}$, $f^{\orto}_{\mu\nu}$ and
$f^{\orto}_{\mu\nu\rho}$. We can now use (\ref{jna21}) and (\ref{jna23})
in the equations (\ref{jna20}) to group the
coefficients into terms proportional to the independent derivatives of
$f_{\mu}$. The obtained equation has the following general structure:
$$
\int_{\cM} d^{p+1}\xi \sqrt{-\gamma} \Big[ X^{\mu\nu\rho}
f^{\orto}_{\mu\nu\rho} + X^{\mu\nu} f^{\orto}_{\mu\nu} + X^{\mu} f_{\mu} +
\nabla_a\left( X^{\mu\nu a} f^{\orto}_{\mu\nu} + X^{\mu ab}
\nabla_b f_{\mu} + X^{\mu a} f_{\mu} \right)  \Big] =0 \, .
$$
Owing to the fact that $f_{\mu}$, $f^{\orto}_{\mu\nu}$ and
$f^{\orto}_{\mu\nu\rho}$ are independent functions on the world sheet, we
deduce that the first three terms must separately vanish. The
resulting equations read:
\begin{subequations} \label{jna24}
\begin{equation} \label{jna24a}
\Pn^{\nu}_{\lambda} \Pn^{\sigma}_{\rho} B^{\mu(\lambda\rho)} =0 \, ,
\end{equation}
\begin{equation} \label{jna24b}
\Pn^{\sigma}_{\nu} \Big[ B^{\mu\nu} - \nabla_a \left( B^{\mu\rho\nu}u^a_{\rho} +
\Pn^{\nu}_{\lambda}B^{\mu\lambda\rho}u^a_{\rho} \right)  \Big] =0 \, ,
\end{equation}
\begin{equation} \label{jna24c}
\nabla_b \left( B^{\mu\nu} u^b_{\nu} + 2 B^{\mu(\lambda\rho)}
u^a_{\lambda}
\nabla_a u^b_{\rho} - \nabla_a B^{\mu(\lambda\rho)} u^a_{\lambda} u^b_{\rho}
\right)
- \left( \Pn^{\rho}_{\sigma} B^{\nu(\lambda\sigma)} + \frac{1}{2}
B^{\nu\lambda\rho} \right) {R^{\mu}}_{\nu\lambda\rho} =0 \, .
\end{equation}
\end{subequations}
This leaves us with the surface integral that vanishes itself:
\begin{equation} \label{jna25}
\int_{\del\cM} d^p\lambda \sqrt{-\h} n_a \left( X^{\mu\nu a}
f^{\orto}_{\mu\nu} + X^{\mu ab}\nabla_b f_{\mu} + X^{\mu a} f_{\mu} \right)
=0 \, .
\end{equation}
The components $f^{\orto}_{\mu\nu}$ and $f_{\mu}$, when  evaluated on the
boundary, are mutually independent, but $\nabla_a f_{\mu}$ is not. This is why
we decompose the $\nabla_a$ derivative into components orthogonal and parallel
to the boundary:
\begin{equation} \label{jna26}
\nabla_a f_{\mu} = n_a \nabla_{\orto} f_{\mu} + v^i_a \nabla_i f_{\mu} \, .
\end{equation}
Here, $\nabla_{\orto}\equiv n^a\nabla_a$, $\nabla_i$ is the total covariant
derivative on $\del\cM$, and $v^a_i$ are the boundary coordinate vectors (see
the appendix for details). Now, $f^{\orto}_{\mu\nu}$, $\nabla_{\orto} f_{\mu}$
and $f_{\mu}$ are mutually independent, and the equation (\ref{jna25}) yields
three sets of boundary conditions:
\begin{subequations} \label{jna27}
\begin{equation} \label{jna27a}
\Pn^{\nu}_{\lambda} B^{\mu(\lambda\rho)} u^a_{\rho} n_a \Big|_{\del\cM} =0 \,
,
\end{equation}
\begin{equation} \label{jna27b}
B^{\mu\lambda\rho} u^a_{\lambda} u^b_{\rho} n_a n_b \Big|_{\del\cM} =0 \, ,
\end{equation}
\begin{equation} \label{jna27c}
\Big[ \nabla_i \left( B^{\mu(\lambda\rho)} u^a_{\lambda} u^b_{\rho} v^i_a
n_b \right) - n_b\Big( B^{\mu\nu} u^b_{\nu}
+ 2 B^{\mu(\lambda\rho)}
u^a_{\lambda} \nabla_a u^b_{\rho} - \nabla_a B^{\mu(\lambda\rho)}
u^a_{\lambda} u^b_{\rho} \Big) \Big] \Big|_{\del\cM} =0 \, .
\end{equation}
\end{subequations}

The equations (\ref{jna24}) and (\ref{jna27}) describe branelike matter
in the pole-dipole approximation. As
we can see, the basic variables $z^{\mu}$, $B^{\mu\nu}$ and
$B^{\mu\nu\rho}$ are mixed in a way that makes it difficult to recognize
their physical meaning. In what follows, we shall decompose the
$B$-coefficients into components orthogonal and parallel to the world sheet,
and try to diagonalize the world sheet equations.

We begin with the $B^{\mu\nu\rho}$ coefficients. Using the constraint
(\ref{jna24a}) to eliminate some orthogonal components, we arrive at
\begin{equation} \label{jna28}
B^{\mu\nu\rho} = 2 u_b^{(\mu} B^{\nu)\rho b}_{\orto} + u_a^{\mu} u_b^{\nu}
B^{\rho ab}_{\orto} + u_a^{\rho} B^{\mu\nu a} \, ,
\end{equation}
where $B^{(\mu\nu)a}_{\orto}\equiv B^{\mu[ab]}_{\orto} \equiv B^{[\mu\nu]a}
\equiv 0$. Note that the $B^{\mu\nu a}$ component is left as is, neither
orthogonal nor parallel to the world sheet. This is because we remember the
extra symmetry~1, which tells us that $B^{\mu\nu a}$ is likely to drop from
the diagonalized world sheet equations. Now, we use (\ref{jna28}) and rewrite
equation (\ref{jna24b}) in the form
\begin{equation} \label{jna29}
\Pn^{\rho}_{\nu} \Big[ B^{\mu\nu} - \nabla_a\left( S^{\mu\nu a} + N^{\mu\nu a}
\right) \Big] =0 \, ,
\end{equation}
where
\begin{equation} \label{jna30}
S^{\mu\nu a} \equiv B^{\mu\nu a}_{\orto} + u_b^{[\mu} B^{\nu] ba}_{\orto} \, ,
\qquad
N^{\mu\nu a} \equiv B^{\mu\nu a} + u_b^{(\mu} B^{\nu) ba}_{\orto}  \, .
\end{equation}
The new coefficients $S^{\mu\nu a}$ and $N^{\mu\nu a}$ are introduced for
later convenience, and are neither orthogonal nor parallel to the world sheet.
Instead, the defining relations (\ref{jna30}) imply the constraint
\begin{equation} \label{jna31}
S^{\mu\nu[a}u^{b]}_{\nu} =0 \, .
\end{equation}
The coefficients $N^{\mu\nu a} = N^{\nu\mu a}$ and $S^{\mu\nu a}= - S^{\nu\mu
a}$, subject to constraint (\ref{jna31}), are in $1-1$ correspondence with
$B^{\mu\nu\rho}$. In what
follows, we shall rewrite the $B^{\mu\nu\rho}$ coefficients in all our
equations in terms of $S^{\mu\nu a}$ and $N^{\mu\nu a}$:
\begin{equation} \label{jna32}
B^{\mu\nu\rho} = 2 u_a^{(\mu} S^{\nu)\rho a} + N^{\mu\nu a} u_a^{\rho} \, .
\end{equation}

Let us now decompose the $B^{\mu\nu}$ coefficients. With the help of the
projectors (\ref{jna22}), we obtain
\begin{subequations} \label{jna33}
\begin{equation} \label{jna33a}
B^{\mu\nu} = B^{\mu\nu}_{\orto} + 2 u_b^{(\mu}B^{\nu) b}_{\orto} + u_a^{\mu}
u_b^{\nu} B^{ab} \, .
\end{equation}
When used in the equation (\ref{jna29}), this decomposition yields
\begin{equation} \label{jna33b}
B^{\mu\nu}_{\orto} = \Pn^{\mu}_{\lambda} \Pn^{\nu}_{\rho} \nabla_a
N^{\lambda\rho a} \, , \qquad
B^{\mu a}_{\orto} = u^a_{\lambda} \Pn^{\mu}_{\rho} \nabla_b \left(
S^{\lambda\rho b} + N^{\lambda\rho b} \right),
\end{equation}
\end{subequations}
and
\begin{subequations} \label{jna34}
\begin{equation} \label{jna34a}
\Pn^{\mu}_{\lambda} \Pn^{\nu}_{\rho} \nabla_a S^{\lambda\rho a} =0 \, .
\end{equation}
The equations (\ref{jna33b}) and (\ref{jna34a}) are equivalent to
(\ref{jna29}). The first shows that $B^{\mu\nu}_{\orto}$ and $B^{\mu
a}_{\orto}$ are fully fixed by $S$ and $N$. This leaves us with $B^{ab}$,
$S^{\mu\nu a}$ and $N^{\mu\nu a}$ as the only independent coefficients in the
theory. The second is viewed as a partial covariant conservation equation of
the world sheet currents $S^{\mu\nu a}$.

Now, we can use (\ref{jna32}) and (\ref{jna33}) to rewrite
the remaining equation (\ref{jna24c}) in terms of the independent
coefficients. By doing so, we arrive at
\begin{equation} \label{jna34b}
\nabla_b \left( m^{ab}u_a^{\mu} - 2 u^b_{\lambda} \nabla_a S^{\mu\lambda
a} + u_c^{\mu} u^c_{\rho} u^b_{\lambda} \nabla_a S^{\rho\lambda a}
\right) - u_a^{\nu} S^{\lambda\rho a} {R^{\mu}}_{\nu\lambda\rho} =0 \, ,
\end{equation}
\end{subequations}
where
\begin{equation} \label{jna35}
m^{ab} \equiv B^{ab} - u^a_{\rho} u^b_{\lambda} \nabla_c N^{\rho\lambda c}  \,
.
\end{equation}
The world sheet tensor $m^{ab}$ is symmetric, and is used instead of $B^{ab}$
in the set of free coefficients. As we can see, the coefficients $N^{\mu\nu
a}$ have dropped from the world sheet equations (\ref{jna34}),
as expected. The physical meaning of the remaining
coefficients, $m^{ab}$ and $S^{\mu\nu a}$, will be clarified in the next
section.

We can now apply the above procedure to the boundary conditions
(\ref{jna27}). Using the algebraic
constraints (\ref{jna32}), (\ref{jna33}) and (\ref{jna35}),
the boundary conditions are rewritten in terms of the independent
coefficients:
\begin{subequations} \label{jna36}
\begin{equation} \label{jna36a}
S^{\mu\nu a} n_a n_{\nu} \Big|_{\del\cM}=0 \, ,
\end{equation}
\begin{equation} \label{jna36b}
\Pn^{\mu}_{\lambda} \Pn^{\nu}_{\rho} S^{\lambda\rho a} n_a \Big|_{\del\cM} =0
\,
,
\end{equation}
\begin{equation} \label{jna36c}
\Big[ \nabla_i \left( N^{ij} v_j^{\mu} + 2 S^{\mu\nu a} n_av^i_{\nu}
\right) - n_b \left( m^{ba}u_a^{\mu} -
2 u^b_{\nu} \nabla_a S^{\mu\nu a} +
u_c^{\mu} u^c_{\rho} u^b_{\lambda} \nabla_a S^{\rho\lambda a}  \right) \Big]
\Big|_{\del\cM} =0 \, ,
\end{equation}
\end{subequations}
where
\begin{equation} \label{jna37}
N^{ij} \equiv N^{\mu\nu a} n_a v^i_{\mu} v^j_{\nu} \, .
\end{equation}
The $N^{ij}$ coefficients are defined on the boundary, and appear nowhere
else.

The equations (\ref{jna34}) and (\ref{jna36}) are the main result of this paper.
They are an equivalent of the covariant conservation equation
(\ref{jna1}) in the pole-dipole approximation, and determine the evolution of the
brane. The free coefficients $m^{ab}$, $S^{\mu\nu a}$ and $N^{ij}$ carry the
information on the internal structure of the brane. In what follows, we shall
analyze the physical meaning of these coefficients, and provide some examples.

\section{\label{InterpretationSection}Physical interpretation}

The free coefficients $m^{ab}$, $S^{\mu\nu a}$ and $N^{ij}$ characterize the
internal structure of the brane. In this section, we shall analyze their
physical meaning and transformation properties.

\subsection{Symmetries}

Let us first derive transformation properties of the free coefficients
$m^{ab}$, $S^{\mu\nu a}$ and $N^{ij}$. To this end, we invert the
decomposition equations (\ref{jna28}), (\ref{jna33a}), and rewrite the
defining relations (\ref{jna30}), (\ref{jna35}) in terms of the original
$B$-coefficients. The transformation properties of the $B$-coefficients have
already been considered in section\ \ref{MultipoleSection}. It has been shown
that $B$'s are tensors with respect to both, spacetime and world sheet
diffeomorphisms. As a consequence,
\begin{itemize}
\item $m^{ab}$, $S^{\mu\nu a}$ and $N^{ij}$ are tensors of the type defined by
their index structure.
\end{itemize}
In particular, $N^{ij}$ is a second rank tensor with respect to the
boundary reparametrizations.

The physical meaning of the $m^{ab}$ coefficients is already known from the
single-pole approximation \cite{Vasilic2006}. It has been shown that $m^{ab}$
represents the covariantly conserved effective $(p+1)$-dimensional
stress-energy
tensor of the brane. In the pole-dipole approximation, its conservation is
violated by the higher order terms.

In addition to diffeomorphisms, two extra symmetries have been discovered in
section\ \ref{MultipoleSection}. The \emph{extra symmetry} 1 is of the algebraic
type, which ensures that only gauge invariant coefficients appear in a
properly diagonalized world sheet equations. Indeed, the transformation laws
(\ref{jna12}) with the constraint (\ref{jna13}) straightforwardly lead to:
\begin{equation} \label{jna38}
\delta_1 m^{ab}=0 \, , \qquad \delta_1 S^{\mu\nu a} =0 \, , \qquad \delta_1
N^{ij}=0 \, .
\end{equation}

The appearance of the peculiar $N^{ij}$ coefficients that live exclusively on
the boundary is a consequence of the constraint (\ref{jna13}) that parameters
of the extra symmetry 1 obey on the boundary. If not for this, the
transformation law $\delta_1 N^{\mu\nu a} = -\epsilon^{\mu\nu a}$ would imply
that $N^{\mu\nu a}$ are pure gauge \emph{everywhere}, and would have to
disappear from the gauge invariant world sheet equations. Physically, the
$N^{ij}$ coefficients characterize the tangential component of the brane
thickness. Namely, when an infinitely thin brane is thickened, this is done in
all spatial directions. Obviously, thickening in the directions tangential to
the brane surface changes nothing in the brane interior. This is because
matter is not localized along these directions anyway. However, if the brane
is open, the tangential thickening does influence the brane boundary. The
boundary structure thus obtained is characterized by the $N^{ij}$
coefficients. In fact, $N^{ij}$ is a correction to the
effective $p$-dimensional stress-energy tensor of the brane boundary, very
much like $m^{ab}$ is $(p+1)$-dimensional effective stress-energy tensor of
the brane itself. The best way to see this is to consider a brane with extra
massive matter attached to its boundary. The procedure of
section\ \ref{EquationsSection} then yields the generalized boundary conditions
in which the $N^{ij}$ term appears as a correction to the effective boundary
stress-energy tensor $m^{ij}$. An example of the kind is considered in the
next section. It consists of the spinless string with massive, spinning
particles attached to its ends.

The \emph{extra symmetry} 2 has been defined in section\ \ref{MultipoleSection}
as the symmetry generated by the change of the surface $x^{\mu}=z^{\mu}(\xi)$
used in the $\delta$-function expansion (\ref{jna2}). The transformation laws
(\ref{jna14}), (\ref{jna18}), thus obtained, can be used in the derivation of
the corresponding transformation properties of the coefficients $m^{ab}$,
$S^{\mu\nu a}$ and $N^{ij}$. We shall first decompose the parameters
$\epsilon^{\mu}$ into components orthogonal and parallel to the world sheet:
\begin{equation} \label{jna39}
\epsilon^{\mu} = \epsilon^{\mu}_{\orto} + u_a^{\mu} \epsilon^a \,  .
\end{equation}
Then, the direct calculation yields
\begin{subequations} \label{jna40}
\begin{equation} \label{jna40a}
\delta_2 m^{ab} = - \left( u^c_{\mu} m^{ab} + u^{(a}_{\mu} m^{b)c} \right)
\nabla_c \epsilon^{\mu}_{\orto}
+ \left( \epsilon^c\nabla_c m^{ab} - m^{bc} \nabla_c \epsilon^a - m^{ac}
\nabla_c \epsilon^b \right) \,  ,
\end{equation}
\begin{equation} \label{jna40b}
\delta_2 S^{\mu\nu a} = - m^{ab} u_b^{[\mu}\epsilon^{\nu]}_{\orto} \, ,
\end{equation}
\begin{equation} \label{jna40c}
\delta_2 N^{ij} = - m^{ab} v^i_a v^j_b \epsilon^c n_c \, ,
\end{equation}
and, of course,
\begin{equation} \label{jna40d}
\delta_2 z^{\mu} = \epsilon^{\mu}_{\orto} + u_a^{\mu}\epsilon^a  \, .
\end{equation}
\end{subequations}
This transformation rule leaves the world sheet equations (\ref{jna34}) and
(\ref{jna36}) invariant. Notice, however,
that the tangential parameters $\epsilon^a$ do not define a fully independent
symmetry. This is because the subgroup defined by $\epsilon^{\mu}_{\orto}=0$
and $\left.\epsilon^a n_a \right|_{\del\cM}=0$ coincides with the world sheet
reparametrizations $\xi^{a'}=\xi^a + \epsilon^a(\xi)$. This is easily seen if
we remember that $S^{\mu\nu a}$, $N^{ij}$ and $\epsilon^{\mu}$ are of the
order $\cO_1$, and that $\cO_2$ terms are ignored in the pole-dipole
approximation. The parameters $\epsilon^a$ which do not satisfy the boundary
condition $\left.\epsilon^a n_a \right|_{\del\cM}=0$ cannot be associated
with reparametrizations. This is why, in general, we cannot get rid of the
$\epsilon^a$ part of the extra symmetry 2.

The transformation laws (\ref{jna40}) are used for fixing
the gauge freedom of the world sheet equations. As explained in
section\ \ref{MultipoleSection}, the gauge fixing of the extra symmetry 2
corresponds
to the choice of the \emph{central surface of mass} --- the surface that
approximates a branelike matter distribution. In the particle case, it
coincides with the usual notion of the centre of mass. We shall see later how
an appropriate gauge fixing ensures that particle trajectories in flat
spacetimes coincide with straight lines.

\subsection{Intrinsic angular momentum}

There are several ways one can associate the $S^{\mu\nu a}$ coefficients with
the intrinsic angular momentum of the brane. One is to compare the $0$-brane
equations (\ref{jna34}) with the Papapetrou result for the
particle trajectory in the pole-dipole approximation \cite{Papapetrou1951}.
Another is the direct calculation of the angular momentum tensor
$M^{\mu\nu\rho} \equiv T^{\rho[\mu}x^{\nu]}$. In this section, however, we
shall simply count the number of independent charges associated with the
$(p+1)$-currents $S^{\mu\nu a}$.

Let us, first, choose an appropriate coordinate system. To this end, we pick
an arbitrary point of the brane, and attach inertial spacetime and world sheet
frames to it. With this, $g_{\mu\nu}$ and $\gamma_{ab}$ in the chosen point
reduce to $\eta_{\mu\nu}$ and $\eta_{ab}$, respectively. Then, an additional
Lorentz rotation of the spacetime frame is performed to ensure its comoving
character:
$$
u_a^{\mu} = \delta_a^{\mu} \, .
$$
In this gauge, the algebraic constraint (\ref{jna31}) reduces to
\begin{subequations} \label{jna41}
\begin{equation} \label{jna41a}
S^{\mu ab} = S^{\mu ba} \, .
\end{equation}

Now, we count the number of independent charge densities $S^{\mu\nu 0}$. In
the first step, we use the constraint (\ref{jna41a}), and the antisymmetry
condition
\begin{equation} \label{jna41b}
S^{\mu\nu a} = - S^{\nu\mu a}
\end{equation}
\end{subequations}
to rule out the vanishing $S^{abc}$ coefficients. This leaves us with the
charge densities $S^{\bar{\mu}\bar{\nu}0}$ and $S^{\bar{\mu}a0}$. (Here, the
index decomposition $\mu = (a,\bar{\mu})$ is used.) As $\bar{\mu}$ takes
$D-p-1$ values, there are $(D-p-1)(D-p-2)/2$ independent
$S^{\bar{\mu}\bar{\nu}0}$ coefficients, and $(D-p-1)(p+1)$ independent
$S^{\bar{\mu}a0}$ coefficients. In total, there are
\begin{subequations} \label{jna42}
\begin{equation} \label{jna42a}
\ds\frac{D(D-1)}{2} - \frac{(p+1)p}{2} \equiv
\dim \left[ SO(1,D-1) \right] - \dim \left[ SO(1,p) \right]
\end{equation}
independent charge densities $S^{\mu\nu 0}$.

As we can see, the number of independent charges associated with the currents
$S^{\mu\nu a}$ is given as a difference of two terms. The first coincides with
the dimension of $SO(1,D-1)$ group, or equivalently, the number of independent
Lorentz rotations in $D$ spacetime dimensions. The second is the dimension of
$SO(1,p)$ group, i.e. the number of independent Lorentz rotations in
$(p+1)$-dimensional world sheet. Thus, our charges correspond to the Lorentz
rotations perpendicular to the world sheet. Naturally, we associate them with
the intrinsic angular momentum of the brane.

Notice that among the charges $S^{\mu\nu 0}$ there are none
corresponding to the tangential world sheet rotations. This is because they
are already taken into account through the effective stress-energy tensor
$m^{ab}$ of the brane. Indeed, these rotations do not require a nontrivial
brane thickness --- they exist already in the single-pole approximation. In
contrast, the possibility to have perpendicular rotations in the
\emph{comoving frame} demands a thick brane, as simulated by the pole-dipole
approximation. The $S^{\mu\nu a}$ coefficients "measure" the brane thickness,
and have nothing to do with infinitely thin branes. As a consequence, the
angular momentum components associated with the $p(p+1)/2$ tangential
rotations are related to $m^{ab}$ rather than $S^{\mu\nu a}$.

In what follows, we shall use the notation $a=(0,\bar{a})$ to further
decompose the $S^{\mu\nu a}$ coefficients. Thus, the nonvanishing charge
densities are written as $S^{\bar{\mu}\bar{\nu}0}$, $S^{\bar{\mu}\bar{a}0}$
and $S^{\bar{\mu}00}$. They correspond to the $\bar{\mu}-\bar{\nu}$,
$\bar{\mu}-\bar{a}$ and $\bar{\mu}-0$ rotation planes, respectively. The
$S^{\bar{\mu}\bar{\nu}0}$ and $S^{\bar{\mu}\bar{a}0}$ are the spatial angular
momentum components, and $S^{\bar{\mu}00}$ are boosts.

Now, we can use the gauge freedom of extra symmetry 2 to fix some unphysical
coefficients. To this end, the transformation law (\ref{jna40b}) is rewritten
in the comoving frame $u_a^{\mu}=\delta_a^{\mu}$, and applied to the boosts
$S^{\bar{\mu}00}$. The resulting rule
$$
\delta_2 S^{\bar{\mu}00} = \frac{1}{2} m^{00}\epsilon^{\bar{\mu}}
$$
shows that the boosts $S^{\bar{\mu}00}$ are pure gauge, and can be gauged
away. Thus, we are left with the spatial angular momentum densities
$S^{\bar{\mu}\bar{\nu}0}$ and $S^{\bar{\mu}\bar{a}0}$ as the only physical
charge densities associated with the currents $S^{\mu\nu a}$. By direct
counting, we find that there are precisely
\begin{equation} \label{jna42b}
\ds \frac{(D-1)(D-2)}{2} - \frac{p(p-1)}{2} \equiv
\dim \left[ SO(D-1) \right] - \dim \left[ SO(p) \right]
\end{equation}
\end{subequations}
independent charges. They correspond to the \emph{spatial rotations
perpendicular to the brane}. In what follows, the intrinsic angular momentum
of the brane will be referred to as classical spin, or simply spin, for
short. It should not be confused with the usual notion of spin, which
originates from the nonvanishing spin-tensor.

By inspecting the world sheet equations (\ref{jna34}), we see that the
currents $S^{\mu\nu a}$ are coupled to both, the spacetime curvature, and the
brane orbital degrees of freedom. It is only in the particle case that the
spin-orbit interaction can be gauged away. This is done by the proper
definition of the particle centre of mass. As a consequence, the particle
trajectories in flat spacetime coincide with straight lines.

In what follows, we shall consider some examples to demonstrate the influence
of classical spin on the brane dynamics and conserved quantities.

\section{\label{ExamplesSection}Examples}

In this section, the $p=0$ and $p=1$ branes are considered in $4$ spacetime
dimensions. Let us first analyze the general particle case.

\subsection{Particle}

The world sheet of a particle is one-dimensional, and is called world line. We
shall parametrize it with the proper distance $s$, thereby fixing the
reparametrization invariance:
$$
\gamma = u^{\mu}u_{\mu} =-1 \, .
$$
Here, and in what follows, the indices $a,b,\dots$ are omitted, as they take
only one value. Thus, the world line equations (\ref{jna34}) are rewritten as
\begin{subequations} \label{jna43}
\begin{equation} \label{jna43a}
\Pn^{\mu}_{\lambda} \Pn^{\nu}_{\rho} \frac{DS^{\lambda\rho}}{ds}=0 \, ,
\end{equation}
\begin{equation} \label{jna43b}
\frac{D}{ds}\left( mu^{\mu} + 2u_{\nu}\frac{DS^{\mu\nu}}{ds} \right) -
u^{\nu}S^{\lambda\rho}{R^{\mu}}_{\nu\lambda\rho}=0 \, ,
\end{equation}
\end{subequations}
where $Dv^{\mu}/ds \equiv dv^{\mu}/ds +
{\itGamma^{\mu}}_{\lambda\rho}u^{\lambda}v^{\rho}$ \, .
These equations are the same as obtained by Papapetrou
\cite{Papapetrou1951}. The coefficients
$S^{\mu\nu}$ are antisymmetric, but otherwise arbitrary (the constraint
(\ref{jna31}) is identically satisfied in the $p=0$ case). We can, still,
use the gauge freedom of the extra symmetry 2 to
fix the $S^{\mu\nu}u_{\nu}$ components. Indeed, the transformation law
(\ref{jna40b}) implies
$$
\delta_2 (S^{\mu\nu}u_{\nu}) = \frac{m}{2} \epsilon^{\mu}_{\orto},
$$
wherefrom we see that parallel components of $S^{\mu\nu}$ can be gauged away.
This leaves us with
\begin{equation} \label{jna44}
S^{\mu\nu} = S^{\mu\nu}_{\orto} \, .
\end{equation}
The fact that $S^{\mu\nu}_{\orto}$ is orthogonal to the world line is used in
the derivation of the conservation laws. First, we project (\ref{jna43b}) onto
$u_{\mu}$, and obtain
\begin{equation} \label{jna45}
\frac{Dm}{ds}= \frac{dm}{ds} =0 \, .
\end{equation}
Thus, the mass parameter $m$ is conserved along the world line. As a
consequence, the equation (\ref{jna43b}) implies
$$
\frac{D u^{\mu}}{ds} = \cO_1 \, .
$$
Using this, and the fact that $\cO_2$ terms are discarded in the pole-dipole
approximation, the world line equations (\ref{jna43}) are
rewritten as
\begin{subequations} \label{jna46}
\begin{equation} \label{jna46a}
\frac{DS^{\mu\nu}_{\orto}}{ds}=0 \, ,
\end{equation}
\begin{equation} \label{jna46b}
m\frac{D u^{\mu}}{ds} = {R^{\mu}}_{\nu\lambda\rho} S^{\lambda\rho}_{\orto}
u^{\nu} \, .
\end{equation}
\end{subequations}
As we can see, the intrinsic angular momentum $S^{\mu\nu}_{\orto}$ is
covariantly conserved, and measures geodesic deviation of the particle
trajectory.

Finally, let us mention that the boundary conditions (\ref{jna36}) are
absent in the $p=0$ case.

\subsection{String}

The string trajectory is a two-dimensional world sheet with one-dimensional
boundary. As in the particle case, the boundary line will be parametrized with
the proper distance $s$, and the indices $i,j,\dots$, which take only one
value,
will be omitted. Thus, the boundary metric $\h$, and the tangent vector $v^a$
satisfy
$$
\h = v^av_a =-1 \, .
$$

The only peculiarity of the string dynamics, as compared to higher branes, is
the
possibility to gauge away the $N^{ij}$ coefficients. Indeed, there is only one
such
component in the string case, and one free parameter in the transformation law
(\ref{jna40c}):
$$
\delta_2 N = - m^{ab}v_av_b \epsilon \, ,
$$
where $\epsilon\equiv \epsilon^a n_a$. Thus, one can fix the gauge $N=0$,
whereupon the parameters $\epsilon^a$ are constrained to obey $\left.
\epsilon^an_a\right|_{\del\cM}=0$. With this condition, the $\epsilon^a$ part
of the extra symmetry~2 reduces to the reparametrizations.

In what follows, we shall describe two specific string configurations. The
first is
a massive rod rotating around its longitudinal axis. The second is a spinless
Nambu-Goto string with massive spinning particles attached to its ends.

\paragraph{Spinning rod.} In this example, a massive rod slowly spinning
around
its longitudinal axis is considered. For simplicity, we choose flat spacetime
(${R^{\mu}}_{\nu\lambda\rho}=0$), and Cartesian coordinates
($g_{\mu\nu}(x)=\eta_{\mu\nu}$).

The simple solution we shall look for is described as follows. The rod is at
rest,
and lies along the $x$-axis between the points $x=L/2$ and $x=-L/2$. It
rotates
around its longitudinal axis, so that
$$
S^a \equiv S^{23a} = -S^{32a}
$$
are the only nonvanishing $S^{\mu\nu a}$ currents.
The world sheet coordinates $\xi^a$ are fixed by the reparametrization gauge
$z^a(\xi)=\xi^a$, while the boundary parameter $\lambda$ coincides with the
proper
length $s$. As a consequence,
$$
u_a^{\mu}=\delta_a^{\mu} \, , \qquad v^a=\delta_0^a\, ,\qquad
\gamma_{ab}=\eta_{ab}\, ,\qquad \h=-1\, .
$$
One easily verifies that this is a solution of the world sheet equations
(\ref{jna34})
and the boundary conditions (\ref{jna36}), provided
\begin{subequations} \label{jna47}
\begin{equation} \label{jna47a}
\del_a m^{ab} =0 \, , \qquad \del_a S^a =0\, ,
\end{equation}
and
\begin{equation} \label{jna47b}
m^{a1}(\xi^1=\pm {\ts \frac{L}{2}})=0\, ,\qquad S^1(\xi^1=\pm {\ts
\frac{L}{2}}) =0\, .
\end{equation}
\end{subequations}
The equations (\ref{jna47a}) tell us that the effective stress-energy tensor
$m^{ab}$ and
the angular momentum current $S^a$ are conserved quantities. The equations
(\ref{jna47b})
state that there is no flow of energy, momentum and angular momentum through
the boundary.

The only thing that might not be obvious in this example is that the rod is
indeed spinning around its longitudinal axis. To check this, we calculate the
total
angular momentum
\begin{equation} \label{jna48}
J^{\mu\nu} \equiv \int d^3x \; x^{[\mu}T^{\nu]0}\, ,
\end{equation}
and find
$$
J^{23}=\int_{-\frac{L}{2}}^{\frac{L}{2}} dx S^0(t,x) \, ,\qquad
J^{12}=J^{13}=0 \, .
$$
Thus, the rod is indeed rotating in the $y-z$ plane. At the same time, the
energy of the rod,
as given by
\begin{equation} \label{jna49energija}
E=\int d^3x\; T^{00}\, ,
\end{equation}
is shown to coincide with the rod mass:
$$
E=\int_{-\frac{L}{2}}^{\frac{L}{2}} dx\; m^{00}(t,x)\, .
$$
The absence of the kinetic term due to rotation is a consequence of the
adopted approximation.
Indeed, the rotational energy is quadratic in $S^a$, which gives the
negligible
$\cO_2$ contribution to the total energy.

Let us notice, in the end, that both the angular momentum $\vec{J}$ and the
energy $E$ are
conserved during evolution. This follows immediately from the world sheet
equations
(\ref{jna47a}) and boundary conditions (\ref{jna47b}).

\paragraph{Generalized Nambu-Goto string.} In this example, we shall consider
a spin\-less
Nam\-bu-Go\-to string with massive spinning particles attached to its ends.
The stress-energy
tensor is written as a sum of two terms,
\begin{subequations} \label{jna49}
\begin{equation} \label{jna49a}
T^{\mu\nu} = T^{\mu\nu}_{\rm s} + T^{\mu\nu}_{\rm p}\, ,
\end{equation}
where
\begin{equation} \label{jna49b}
T^{\mu\nu}_{\rm s} = \int_{\cM} d^2\xi \sqrt{-\gamma}
B_{\rm s}^{\mu\nu} \frac{\delta^{(4)}(x-z)}{\sqrt{-g}} \, ,
\end{equation}
\begin{equation} \label{jna49c}
T^{\mu\nu}_{\rm p} = \int_{\del\cM} d\lambda \sqrt{-\h} \left(
B_{\rm p}^{\mu\nu}\frac{\delta^{(4)}(x-z)}{\sqrt{-g}} - \nabla_{\rho}
B_{\rm p}^{\mu\nu\rho} \frac{\delta^{(4)}(x-z)}{\sqrt{-g}} \right) .
\end{equation}
\end{subequations}
The string part of the stress-energy tensor is written in the single-pole
approximation, in
accordance with the assumed absence of spin in the string interior. The usual
procedure then
yields the familiar world sheet equations
\begin{equation} \label{jna50}
\nabla_a \left( m^{ab} u_b^{\mu} \right) =0\, .
\end{equation}
The particle part $T^{\mu\nu}_{\rm p}$ has the general form (\ref{jna11}),
constrained by the
requirement that particle trajectories coincide with the string boundary. The
resulting
boundary conditions are interpreted as the particle equations of motion:
\begin{subequations} \label{jna51}
\begin{equation} \label{jna51a}
\Pnn^{\mu}_{\lambda} \, \Pnn^{\nu}_{\rho} \frac{D S^{\lambda\rho}}{ds} =0\, ,
\end{equation}
\begin{equation} \label{jna51b}
\frac{D}{ds} \left( mv^{\mu} +2v_{\nu} \frac{D S^{\mu\nu}}{ds} \right)
- v^{\nu}S^{\lambda\rho} {R^{\mu}}_{\nu\lambda\rho} = n_a m^{ab} u_b^{\mu} \, .
\end{equation}
\end{subequations}
Here, $\Pnn^{\mu}_{\nu} \equiv \delta^{\mu}_{\nu}+v^{\mu}v_{\nu}$ is the
orthogonal projector
to the string boundary, and should not be confused with $\Pn^{\mu}_{\nu}$. The
boundary
conditions (\ref{jna51}) differ from the particle world line equations
(\ref{jna43})
by the presence of the string force on the right-hand side. As the boundary
$\del\cM$
consists of two disjoint lines, the mass and spin of the two particles may
differ.

In what follows, we shall assume that the string is made of the Nambu-Goto
type
of matter, moving in a $4$-dimensional flat spacetime:
$$
m^{ab} = T \gamma^{ab}\, , \qquad {R^{\mu}}_{\nu\lambda\rho}=0\, .
$$
Then, the world sheet equations (\ref{jna50}) reduce to the familiar
Nambu-Goto
equations, and the string force on the right-hand side of (\ref{jna51b})
becomes
$T n^{\mu}$. As for the particles, we shall impose the constraint
\begin{equation} \label{jna52}
S^{\mu\nu}v_{\nu}=0\, ,
\end{equation}
which rules out the boost degrees of freedom. Physically, this condition
constrains the particle centre of mass to coincide with the string end,
with accuracy $\cO_2$. After this, we are left with
$$
\vec{S}\equiv \lc^{0\lambda\rho\mu} S_{\lambda\rho} \vec{e}_{\mu}
$$
as the only independent components of $S^{\mu\nu}$.

Now, we look for a simple, \emph{straight line} solution of the equations of
motion (\ref{jna50}). Without loss of generality, we put
$$
\vec{z} = \vec{\alpha} (\tau) \sigma\, ,\qquad z^0=\tau\, ,
$$
where $\xi^0\equiv \tau$ and $\xi^1\equiv \sigma$ take values in the intervals
$(-\infty,\infty)$ and $[-1,1]$, respectively. Assuming that the string length
$L=2|\vec{\alpha}|$, and the velocity of the string ends
$V=|d\vec{\alpha}/d\tau|$ are constant, the equation (\ref{jna50})
reduces to
$$
\frac{d^2}{d\tau^2}\vec{\alpha} + \omega^2\vec{\alpha}=0 \, , \qquad
\omega\equiv \frac{2V}{L} \, .
$$
It describes uniform rotation in a plane. Choosing the rotation plane to be
the $x-y$ plane, we get the solution
\begin{equation} \label{jna53}
\vec{\alpha} = \frac{L}{2} \left( \cos \omega\tau \,\vec{e}_x +
\sin \omega\tau \,\vec{e}_y \right) .
\end{equation}

Next we consider the boundary equations (\ref{jna51}). Omitting the details of
the calculation, we find that the particle intrinsic angular momentum
satisfies
\begin{equation} \label{jna54}
\frac{d \vec{S}}{d\tau}=0  \, , \qquad \vec{S}=S\vec{e}_z \, ,
\end{equation}
while its velocity becomes
\begin{equation} \label{jna55}
V = \frac{1}{\sqrt{1+\frac{2\mu}{TL}}} \, ,\qquad \mu \equiv m +
\sqrt{\frac{2T}{mL}}
S \, .
\end{equation}
Each of the two particles has its own mass and intrinsic angular momentum,
denoted by $m_{\pm}$ and $S_{\pm}$ for the particle at $\sigma=\pm 1$. As
both particles have the same velocity, their masses are related by
$\mu_+=\mu_-$. We see that the particle masses $m_{\pm}$ may differ, in
spite of the fact that the centre of mass of the string-particle system
is at $\sigma=0$. This is a consequence of the nontrivial spin-orbit
interaction that contributes to the total energy.

By inspecting the expression (\ref{jna55}), we see that $V<1$, as it should
be.
In the limit $\mu\to 0$, the string ends move with the speed of light,
representing
the Nambu-Goto dynamics with Neumann boundary conditions. When $\mu\to\infty$,
the string ends do not move. This is an example of Dirichlet boundary
conditions.

The total angular momentum and energy of the considered system are calculated
using (\ref{jna48}) and (\ref{jna49energija}). One finds
$$
E=TL\frac{\arcsin V}{V} + \frac{2\mu}{\sqrt{1-V^2}} - \frac{2V}{L}\left(
S_++S_- \right) ,
$$
$$
J = \frac{TL^2}{4} \left( \frac{\arcsin V}{V^2} - \frac{\sqrt{1-V^2}}{V}
\right)
+ \frac{2\mu}{\sqrt{1-V^2}} \frac{LV}{2} + S_++S_- \, .
$$
These equations have obvious interpretation. The total energy of the system
consists of the string energy, kinetic energy of the two particles, and the
spin-orbit interaction energy. The particle intrinsic rotational energy, being
quadratic in $\vec{S}$, is neglected in the pole-dipole approximation.
 The total angular momentum
includes the orbital angular momentum of the string and the two particles,
and the particle spins.

In the limit of small particle masses, the free
parameter $L$ can be eliminated in favour of $E$, which leads to
$$
J = \frac{1}{2\pi T}E^2 + 2\left( S_++S_- \right) .
$$
The first term on the right-hand side defines the known Regge trajectory,
while
the second represents a small correction due to the presence of spinning
particles at the string ends. As we can see, the unique Regge trajectory of
the
ordinary string theory splits into a family of distinctive trajectories.

\section{\label{ConclusionSection}Concluding remarks}

The study in the preceding sections concerns the dynamics of classical
brane-like matter in curved backgrounds. In the simple case we have
considered,
the target space geometry is Riemannian. The type of matter fields is not
specified. We only assume that matter fields are sharply localized around a
brane.

The method we use is a generalization of the Mathisson-Papapetrou method for
pointlike matter \cite{Mathisson1937,Papapetrou1951}. It has already been used
in \cite{Vasilic2006} for the study of infinitely thin string. In this work,
higher branes are considered in the approximation where nonzero thickness of
the
brane is taken into account. As a consequence, additional degrees of freedom
appear to characterize the intrinsic angular momentum of the brane.

The results of our analysis can be summarized as follows. In
section\ \ref{MultipoleSection} we have refined the Mathisson-Papapetrou method by
developing a manifestly covariant decomposition of the stress-energy tensor
into a series of $\delta$-function derivatives. The truncation of the series at
any level has been proven invariant with respect to both, spacetime and world
sheet diffeomorphisms. We have also utilized two extra gauge symmetries. In
particular, the extra symmetry 2 has been used to properly define the central
surface of branelike mass distribution.

In section\ \ref{EquationsSection} the $p$-brane world sheet equations and
boundary
conditions have been derived in the pole-dipole approximation. Beside the
effective
stress-energy tensor of the brane, a new set of coefficients appear to
characterize
the nonzero brane thickness. They have been interpreted in
section\ \ref{InterpretationSection} as effective brane currents associated with the
intrinsic
angular momentum of the brane. By the proper definition of the central surface
of
mass, we have shown that charges associated with boosts can be gauged away.

Finally, we provided some examples. A particularly interesting one is a
spinless
string with spinning particles attached to its ends. When applied to the
Nambu-Goto
matter, it gives the correction to the behavior of the known Regge
trajectories.

Let us say, in the end, that these results can be generalized to include the
effects
of the background torsion. The brane dynamics in the Riemann-Cartan spacetimes
will
be the objective of our next paper.

\acknowledgments
This work was supported by the Serbian Science Foundation, Serbia.

\appendix

\section{Differential geometry of surfaces}

Throughout the paper we deal with the geometry of surfaces embedded in a
general Riemannian spacetime. Here we introduce some basic notions needed
for the exposition.

Consider a $D$-dimensional Riemannian spacetime parametrized by the coordinates
$x^{\mu}$. The metric tensor
is denoted by $g_{\mu\nu}(x)$ and has Minkowski signature
$\diag(-,+,\dots,+)$. Given the metric, one introduces the Levi-Civita
connection
$$
{\itGamma^{\mu}}_{\rho\sigma} \equiv \frac{1}{2}g^{\mu\lambda} \left(
\del_{\rho}g_{\lambda\sigma} + \del_{\sigma} g_{\lambda\rho} - \del_{\lambda}
g_{\rho\sigma} \right) \, ,
$$
and the covariant derivative $\nabla_{\lambda}$:
$$
\nabla_{\lambda} V^{\mu} \equiv \del_{\lambda} V^{\mu} +
{\itGamma^{\mu}}_{\rho\lambda}V^{\rho} \, .
$$
The Riemann curvature tensor is defined as
$$
{R^{\mu}}_{\lambda\nu\rho} \equiv {\itGamma^{\mu}}_{\lambda\rho,\nu} -
{\itGamma^{\mu}}_{\lambda\nu,\rho} + {\itGamma^{\mu}}_{\sigma\nu}
{\itGamma^{\sigma}}_{\lambda\rho} - {\itGamma^{\mu}}_{\sigma\rho}
{\itGamma^{\sigma}}_{\lambda\nu} \, .
$$

Now introduce a $(p+1)$-dimensional surface $\cM$, parametrized by the coordinates
$\xi^a$. If the surface equation is $x^{\mu}=z^{\mu}(\xi)$, one introduces the
coordinate vectors
$$
u_a^{\mu} \equiv \frac{\del z^{\mu}}{\del\xi^a} \, .
$$
The induced metric tensor on the surface is defined by
$$
\gamma_{ab} = g_{\mu\nu}(z) u_a^{\mu} u_b^{\nu} \, .
$$
Assume that the surface is everywhere regular and
the coordinates well defined. The induced metric is nondegenerate and of
Minkowski signature $\diag(-,+,\dots,+)$.

Given an arbitrary spacetime vector $V^{\mu}$, one can uniquely split it into
a sum of vectors orthogonal and tangential to the surface $\cM$, $V^{\mu} =
V^{\mu}_{\orto} + V^{\mu}_{\para}$, using the projectors
$$
\Pp^{\mu}_{\nu} \equiv u_a^{\mu}u^a_{\nu} \, , \qquad \Pn^{\mu}_{\nu} \equiv
\delta^{\mu}_{\nu} - u_a^{\mu}u^a_{\nu} \, ,
$$
so that $V^{\mu}_{\orto}=\Pn^{\mu}_{\nu}V^{\nu}$ and $V^{\mu}_{\para} =
\Pp^{\mu}_{\nu}V^{\nu}$. Next, one can define the induced connection
${\itGamma^a}_{bc}$ by parallel transporting a vector using the spacetime
connection, and then projecting the result onto the surface $\cM$. If the
connection is defined this way, one can show that it is precisely the
Levi-Civita connection
$$
{\itGamma^{a}}_{bc} = \frac{1}{2}\gamma^{ad}\left( \del_b\gamma_{dc} +
\del_c\gamma_{db} - \del_d\gamma_{bc} \right) .
$$
Now, one can define the \emph{total} covariant derivative $\nabla_a$, that
acts on both types of indices:
$$
\nabla_a V^{\mu b} = \del_a V^{\mu b} +
{\itGamma^{\mu}}_{\lambda\rho}u_a^{\rho} V^{\lambda b} +
{\itGamma^b}_{ca}V^{\mu c} \, .
$$
The metricity conditions $\nabla_a g_{\mu\nu}=\nabla_a \gamma_{bc}=0$ are
identically satisfied. One can also introduce the second fundamental form,
$$
K^{\mu}_{ab} \equiv \nabla_a u_b^{\mu} \, ,
$$
which satisfies the useful identities:
$$
K^{\mu}_{ab}= K^{\mu}_{ba}, \qquad K^{\mu}_{ab}u^c_{\mu}=0 \, .
$$

The surface $\cM$ may have a boundary $\del\cM$, and we denote its coordinates
by $\lambda^i$. The boundary is supposed to satisfy
the analogous geometric requirements as the surface itself. Given the
boundary $\xi^a = \zeta^a(\lambda)$, one introduces its coordinate vectors
$$
v^a_i \equiv \frac{\del\zeta^a}{\del\lambda^i} \, ,
$$
and the induced metric
$$
\h_{ij} = \gamma_{ab}(\zeta) v^a_i v^b_j \, .
$$
The induced connection ${\itGamma^i}_{jk}$ is the Levi-Civita connection,
so that the total covariant derivative  $\nabla_i$, which acts as
$$
\nabla_i V^{\mu bj} = \del_i V^{\mu bj} +
{\itGamma^{\mu}}_{\lambda\rho}v_i^{\rho} V^{\lambda bj} +
{\itGamma^b}_{ca} v^a_i V^{\mu cj} + {\itGamma^j}_{ki} V^{\mu bk} \, ,
$$
satisfies the metricity conditions $\nabla_i g_{\mu\nu} =
\nabla_i \gamma_{ab} = \nabla_i \h_{jk} =0$. Here,
$v_i^{\mu} \equiv u_a^{\mu} v^a_i$ are the spacetime components of the
boundary coordinate vectors. The boundary projectors are defined as
$\Ppp^{\mu}_{\nu}\equiv v_i^{\mu}v^i_{\nu}$ and $\Pnn^{\mu}_{\nu}\equiv
\delta^{\mu}_{\nu}- v_i^{\mu}v^i_{\nu}$.

Throughout the paper, the covariant form of the Stokes theorem is used:
$$
\int_{\cM} d^{p+1}\xi \sqrt{-\gamma} \nabla_a V^a = \int_{\del\cM} d^p\lambda
\sqrt{-\h} n_a V^a  \, .
$$
Here, $n_a$ is the normal to the boundary. It is defined as
$$
n_a = \frac{1}{p!} e_{ab_1\dots b_p} e^{i_1\dots i_p} v^{b_1}_{i_1}\dots
v^{b_p}_{i_p} \, ,
$$
where $e_{ab_1\dots b_p}$ and $e^{i_1 \dots i_p}$ are totally
antisymmetric world tensors on the surface and the boundary,
respectively. They are defined using the Levi-Civita symbols
$\lc_{ab_1\dots b_p}$ and $\lc^{i_1\dots i_p}$, and corresponding
metric determinants:
$$
e_{ab_1\dots b_p}(\xi) \equiv \sqrt{-\gamma}\lc_{ab_1\dots b_p} ,\qquad
e^{i_1 \dots i_p}(\lambda) \equiv \frac{1}{\sqrt{-\h}} \lc^{i_1\dots i_p} \, .
$$
The normal $n_a$ is always spacelike, and satisfies the following identities:
$$
n_an^a= 1 \, , \qquad n_av^a_i=0 \, , \qquad \Pn^{\mu}_{\nu} =
\Pnn^{\mu}_{\nu}
-
n^{\mu}n_{\nu} \, ,
$$
where $n^{\mu}\equiv u_a^{\mu}n^a$.

\end{document}